# An UHF RFID Energy-Harvesting System Enhanced by a DC-DC Charge Pump in Silicon-on-Insulator Technology

Danilo De Donno, Luca Catarinucci, and Luciano Tarricone

*Abstract*—An RF-DC converter enhanced by a DC-DC voltage booster in silicon-on-insulator technology for UHF radio frequency identification (RFID) energy harvesting is presented in this letter. When the received RF power level is -14 dBm or higher, the system, fabricated on an FR4 substrate using off-the-shelf low-cost discrete components and connected to a flexible dipole antenna, is able to produce 2.4-V DC voltage to power general-purpose electronic devices. As a simple proof of concept, a device comprising microcontroller, temperature sensor, and EEPROM is considered in this work. The experimental results demonstrate the capability of the system to autonomously perform temperature data logging up to a distance of 5 m from a conventional UHF RFID reader used as an RF energy source.

*Index Terms*—RF energy harvesting, UHF, RFID, DC-DC charge pump, wireless sensor networks.

## I. INTRODUCTION

IN recent years, the growing interest in the UHF radio frequency identification (RFID) technology is leading to very dense reader deployments, such as in airports, warehouses, hospitals, supermarkets. This amount of RF energy – commercial UHF RFID readers transmit up to 4 Watts of equivalent isotropically radiated power (EIRP) – could be potentially harvested and converted into usable power for smart electronic devices, such as environmental, structural, surveillance, and wearable biological sensors [1].

To the best of our knowledge, only few works in the literature focus on UHF RFID sources for powering general-purpose electronic devices. Among these, a fully-passive wireless identification and sensing platform (WISP) is presented in [2], while RF-DC power conversion systems in CMOS technology are proposed in [3] and [4]. Conversely, the majority of works considers different RF bands or directly relies on natural ambient sources (e.g. solar, thermal and kinetic) to harvest energy. A battery-less device for long-range energy scavenging from terrestrial TV broadcasts in the 500–600 MHz frequency band is implemented in [5]. A power management circuit for wireless sensors using scavenged solar energy is presented in [6]. An RF energy-harvesting circuit generating 1 V at 2.2 GHz is described in [7].

In this letter, an RF energy-harvesting system enhanced by a DC-DC charge pump in silicon-on-insulator (SOI) technology is designed and fabricated using off-the-shelf low-cost discrete components. As a simple proof of concept, the developed circuit is used to power a microcontroller unit (MCU) for temperature data logging. The experimental results demonstrate the capability of the system to produce a 2.4-V DC voltage from received RF power levels as low as -14 dBm, thus enabling autonomous operation up to 5 m of distance from a standard UHF Class-1 Generation-2 RFID reader used as an RF energy source. Such a distance corresponds to a three-times improvement over a conventional rectifying approach not adopting the DC-DC boosting technique. The empirical analysis reveals how the proposed idea represents a suitable expedient, applicable to generic RF-DC conversion circuits, to trade off efficiency against sensitivity. This translates into the possibility to easily and rapidly customize the performance of self-powered devices in terms of operating range and duty cycle depending on the application.

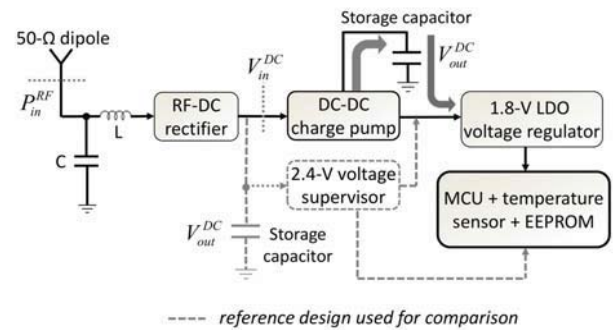

Fig. 1. Block diagram of the proposed RF energy-harvesting system and reference design used for comparison.

## II. SYSTEM DESIGN

The block diagram of the proposed RF energy-harvesting system is shown in Fig. 1. The initial RF-DC converter is a 4-stage voltage multiplier (see [8] for design details) that is matched to a 50-Ω dipole-like antenna by an LC matching network. The Avago HSMS-285C [9] zero-bias Schottky diodes have been chosen as rectifying devices because of their high-detection sensitivity at UHF frequencies. Then, the Seiko Instruments S-882Z24 charge pump IC [10] has been adopted to step-up the rectified voltage $V_{in}^{DC}$. Such a DC-DC converter implements fully depleted SOI technology to enable ultra-low voltage operation. In fact, when $V_{in}^{DC}$ is 0.35 V or higher the oscillation circuit starts to operate and the stepped-up electric power is stored in a storage capacitor (see Fig. 1). When the capacitor reaches $V_{out}^{DC}$=2.4 V, the integrated supervisory circuit of the S-882Z24 automatically releases the stored energy to power a TI MSP430F5310 MCU [11] which samples its temperature sensor and stores the result in an external EEPROM. Note that, unlike a conventional RFID tag [12], the proposed device does not backscatter the logged data. When $V_{out}^{DC}$ decreases to approximately 1.85 V as a result of the storage capacitor discharge, the S-882Z24 disconnects its output and starts a new charging process. The idle time between MCU operations, i.e. the duty cycle of the overall system, is determined by the amount of input power to the charge-pump IC to charge up the storage capacitor. In order to perform a fair comparison, in the reference circuit not implementing the DC-DC booster (dashed elements in Fig. 1) a 2.4-V supervisor is used to provide 0.6 V of headroom on the storage capacitor above the 1.8-V regulated voltage. It was found both theoretically and experimentally that, with the 0.6 V of headroom and 0.5 mA of MCU current consumption for temperature data logging (roughly 9 ms of execution time), a 10-µF storage capacitor is a suitable choice.

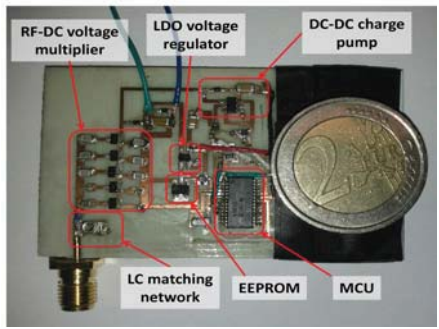

Fig. 2. Prototype photo. Dimensions: 7.5x6 cm$^2$.

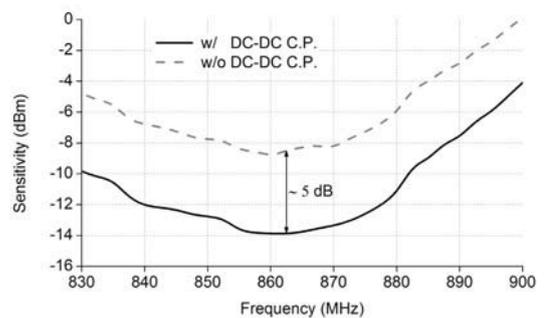

Fig. 4. Over-the-air measurements of the system sensitivity with and without the DC-DC charge pump.

### III. EXPERIMENTAL RESULTS

The designed RF energy-harvesting circuit was fabricated using off-the-shelf discrete components on an FR4 substrate (see Fig. 2 for a prototype photo) and connected via an SMA connector to a 50-Ω dipole-like antenna on a paper substrate. Fig. 3 shows the antenna prototype and the comparison between simulated and VNA-measured return losses. Results confirm a good impedance matching around 866.5 MHz, i.e. the center frequency of the European UHF RFID band.

In a first set of experiments, we measured the sensitivity of the RF energy-harvesting circuit. A VNA was used to inject a continuous 866.5-MHz waveform into the SMA connector of the circuit. Then, the VNA power was iteratively changed and, at each power level, the LC matching network tuned accordingly to achieve the maximum power transfer to the rectifier. The minimum power necessary to activate the DC-DC charge pump, i.e. to charge the storage capacitor, was found to be -14 dBm with L=3.3 nH and C=8.2 pF (see Fig. 3 for the achieved return loss). In order to verify such a result and rapidly perform sensitivity measurements at different frequencies, we conducted an over-the-air analysis in anechoic chamber. A continuous wave (CW) was generated by an RF vector signal generator connected to a circularly polarized antenna (gain $G_{tx}$=5.5 dBi) placed at 1 m of distance from the harvesting device. The minimum transmit power $P_{tx,ON}$ required to start the storage capacitor charge was recorded at different frequencies of the CW in the 830–900 MHz band. Then, the sensitivity value was calculated according to the following equation based on the free-space Friis' propagation model:

$$S = EIRP_{ON} G_{rx} \left(\lambda/4\pi d\right)^2 \eta_{plf} \quad \text{(Watt)} \quad (1)$$

where $EIRP_{ON}=P_{tx,ON}G_{tx}$ is the minimum EIRP required to start the charging process, $G_{rx}$=1.85 dBi is the gain of the dipole-like antenna connected to the harvester, $\lambda$ is the wavelength, $\eta_{plf}$=0.5 is the polarization loss factor due to the circularly polarized transmit antenna, and $d$=1 m is the distance between the RF source and the device under test. In order to evaluate the impact of the DC-DC charge pump over the sole rectifier, the experiments were repeated with the device having the DC-DC booster bypassed and the matching network re-tuned to ensure maximum power transfer (see Fig. 1). The results plotted in Fig. 4 show a sensitivity boost of 5 dB provided by the DC-DC charge pump and corroborate the VNA analysis (peak sensitivity of -14 dBm around 866.5 MHz). In order to explain such a sensitivity improvement, the system efficiencies and achieved DC voltages ($V_{out}^{DC}$) were measured. As shown in Fig. 5, the DC-DC charge pump allows to trade off efficiency against sensitivity. In fact, at the cost of an 11% reduction in peak efficiency, the charge pump enables full operation ($V_{out}^{DC}$=2.4 V) even in *Region 1* where, instead, the conventional approach is not able to generate a sufficient voltage to wake up the MCU.

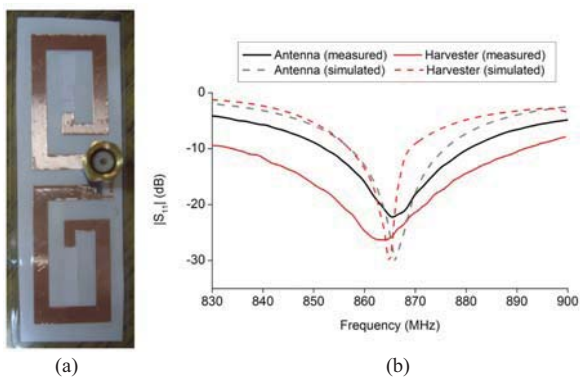

(a)      (b)

Fig. 3. 50-Ω paper-based antenna prototype (a) and comparison between simulated and measured (-14 dBm VNA output power) return losses (b).

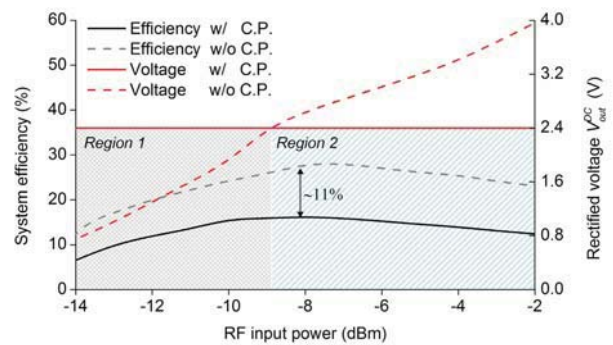

Fig. 5. System efficiency and rectified DC voltages ($V_{out}^{DC}$) with and without the DC-DC charge pump using a VNA for RF signal insertion.

TABLE I. COMPARISON OF POWER HARVESTERS IN THE UHF RFID BAND

| Reference | Frequency [MHz] | Number of stages in the RF-DC voltage multiplier | DC-DC charge pump | Rectifier load | Sensitivity for 2.4-V output voltage | Peak efficiency | Fabrication process |
|---|---|---|---|---|---|---|---|
| This work | 866.5 | 4 | Yes | 3 kΩ | -14 dBm | 16% | discrete components |
| [2] | 915 | 5 | No | 180 kΩ | -8 dBm | 30% | discrete components |
| [3] | 900 | 16 | No | 100 kΩ | -11 dBm | 60% | 0.25 μm CMOS |
| [4] | 920 | 4 | No | 330 kΩ | -8 dBm | n.a. | 0.18 μm CMOS |

In a second set of experiments, the performance of the proposed RF energy-harvesting system was evaluated in real operating conditions. Fig. 6 shows the average temperature logging rate when varying the distance from a commercial UHF RFID reader compliant with the European regulations (3.2 W of maximum EIRP in the 865–868 MHz frequency band). The experiments were conducted in a standard office room with reader antenna and device under test placed along the line of sight (LOS) 1.5 meters above the floor. The results show that, in the 0–1.5 meters range, the delay time needed by the DC-DC charge pump to build up the output voltage causes a lower rate compared to the same RF energy-harvesting system without the DC-DC booster. However, as expected from the sensitivity analysis, the DC-DC charge pump allows to significantly extend the operating range up to 4.8 meters, i.e. approximately three times greater than the range achieved without the DC-DC booster.

The main features of our work compared to other RF energy harvesters in the UHF RFID band are reported in Table I. Our implementation exhibits two main strengths: the first one consists in the adoption of a DC-DC charge pump to further increase the rectified DC voltage; the second one is related to the use of low-cost off-the-shelf discrete components instead of an embedded CMOS-IC design and fabrication [3], [4]. This last point, in conjunction with our precise indication of the required electronic parts, is essential since it allows researchers and practitioners to replicate, customize, and further improve the design here proposed.

IV. CONCLUSION

An RF energy-harvesting system boosted by a DC-DC charge pump in SOI technology has been presented in this work. A prototype featuring temperature sensing and data storage has been fabricated on a printed circuit board using low-cost discrete components and connected to a dipole-like antenna. The system exhibits a sensitivity of -14 dBm and is able to produce a 2.4-V DC voltage when placed at a maximum distance of about 5 m from a commercial UHF RFID reader used as an RF energy source. Such a distance represents a three-times improvement over a conventional rectifying approach not implementing the DC-DC voltage boosting technique.

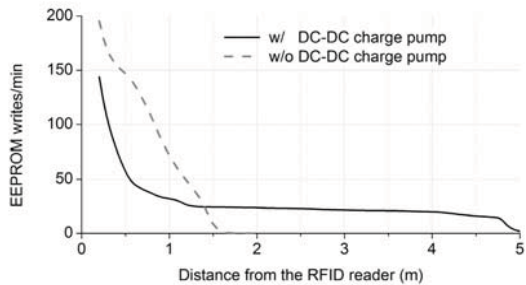

Fig. 6. Temperature logging rate vs. distance from the UHF RFID reader.